\RequirePackage{lineno}
\documentclass[aps,prl,twocolumn,groupedaddress,amsmath,amsfonts]{revtex4-1}
\usepackage{graphicx,siunitx,pgfplots,hyperref}

\hypersetup{
	colorlinks=true,
	linkcolor=blue,
	filecolor=blue,      
	urlcolor=blue,
	citecolor=blue,
}

\newcommand*{\dbul}[2][]{%
	\underline{\underline{ #2}}
}

\newcommand*{\myvec}[1]{
	\mathbf{\underline{#1}}}

\newcommand{\sref}[2]{Fig.\,\hyperref[#1]{\ref{#1}#2}}

\def\spvecA#1;{\if;#1;\else #1\cr \expandafter \spvecA \fi}
\DeclareSIUnit\tei{$\tau_{ei}$}
\DeclareSIUnit\vth{v_T}
\DeclareSIUnit\lmfp{$\lambda_{mfp}$}

\begin{document}  
	
	
	\title{Enhancement of Pressure Perturbations in Ablation due to Kinetic Magnetised Transport Effects under Direct-Drive ICF relevant conditions}
	
	
	\author{D.W. Hill}
	\email[]{dh814@imperial.ac.uk}
	\author{R.J. Kingham}
	\affiliation{Imperial College London}
	
	
	\date{\today}
	
	\begin{abstract}
		
		We present for the first time kinetic 2D Vlasov-Fokker-Planck simulations, including both self-consistent magnetic fields and ablating ion outflow, of a planar ablating foil subject to nonuniform laser irradiation. Even for small hall parameters  ($\omega \tau_{ei} \lesssim 0.05$) self-generated magnetic fields are sufficient to invert and enhance pressure perturbations. The mode inversion is caused by a combination of the Nernst advection of the magnetic field and the Righi-Leduc heat-flux. Non-local effects modify these processes. The mechanism is robust under plasma conditions tested; it is amplitude independent and occurs for a broad spectrum of perturbation wavelengths, $\lambda_p = 10-100\si{\micro\meter}$. The ablating plasma response to a dynamically evolving speckle pattern perturbation, analogous to an optically smoothed beam, is also simulated. Similar to the single mode case, self-generated magnetic fields increase the degree of nonuniformity at the ablation surface by up to an order of magnitude and are found to preferentially enhance lower modes due to the resistive damping of high mode number magnetic fields.  
		
	\end{abstract}
	
	
	\maketitle

Insufficient uniformity of laser irradiation can be a major source of degrading target performance in direct-drive inertial confinement fusion (ICF) \cite{Li2004,Hu2016,Shah2017a}. In direct drive, laser energy is absorbed in hot, low density plasma near the critical surface. The energy must then be transported, predominantly via thermal conduction, towards the interface between the expanding plasma and the solid density target, the ablation surface. Irradiation nonuniformities imprint themselves onto the cold capsule surface during this ablation phase, where they can seed hydrodynamic instabilities, damaging fusion yields. The detrimental effect of nonuniform energy deposition is counteracted via thermal smoothing in the conduction zone, the region between the critical and ablation surfaces, and dynamic overpressure stabilisation at the ablation surface \cite{Nuckolls1972a,Goncharov1996,Sanz1996}.  The conventional view of thermal smoothing is that the electrons transporting energy from the critical surface to the ablation surface conduct some heat sideways during transit. The lateral thermal conduction, according to the cloudy day model \cite{Brueckner1974}, should result in an exponential attenuation of the pressure perturbation amplitudes,  $\delta P/P \propto e^{-kx}$. This picture neglects both kinetic (nonlocal) effects and magnetic fields, both of which can severely alter the magnitudes and directions of heat fluxes \cite{Bell1985a,Braginskii1965}. The temperature scale length within the conduction zone is typically on the order of the electron mean free paths. Under such conditions the electron distribution function deviates from its equilibrium Maxwellian form and the classical (Braginskii) heat transport model \cite{Braginskii1965} breaks down \cite{Malone1975}. 

Experimental measurements have demonstrated that non-local heat transport effects are important in nano-second time scale laser-solid interactions \cite{Gregori2004,Gotchev2006}, and must be taken into account to align ICF simulations with experimental predictions \cite{Michel2015,Igumenshchev2010,Hu2008}. It has been shown that there can be a significant interplay between the non-local heat flux effects and the magnetic field dynamics \cite{Joglekar2014,Ridgers2008a}. 	

Self-generated magnetic fields have been measured in ablation phase ICF experiments \cite{Manuel2012,Igumenshchev2014} and are predicted to be important in a variety of ICF relevant conditions \cite{Walsh2017,Joglekar2014}. Crossed number density, $n_e$, and temperature, $T_e$, gradients, that occur at perturbations in the laser energy deposition, generate magnetic fields through the Biermann battery mechanism, $\partial_t \myvec{B} = -\nabla n_e \times \nabla T_e/(|e|n_e)$ \cite{Stamper1971,Biermann1950}. The Nernst effect \cite{Nishiguchi1984} then advects these fields with the electron heat flux, $q_e$, (at the velocity  $\myvec{v}_N \approx \myvec{q}_e/(\frac{5}{2} n_e T_e)$ \cite{Haines1986}) into the conduction zone and simultaneously, compressively amplifies them.

	
	The magnetised heat transport effect dominant in this study is the Righi-Leduc heat flow, $\myvec{q}_{RL} = -\kappa_{\wedge} (\myvec{b} \times \nabla T_e)$, where $\myvec{b}$ represents the magnetic field unit vector. This is the heat flow, with thermal conductivity $\kappa_{\wedge}$, generated perpendicular to a temperature gradient, $\nabla T_e$, due to the Lorentz force acting upon the heat carrying electrons. In this work, a combination of the Nernst advection and amplification of magnetic fields and the Righi-Leduc heat flow is found to invert and enhance perturbations within the conduction zone.

	A key source of irradiation nonuniformity is irregularity within the laser beams. A variety of beam smoothing techniques (RPP \cite{Kato1984}, SSD \cite{Skupsky1989}, ISI \cite{Lehmberg1987} etc.) are employed on laser systems to mitigate this. The smoothed beams are composed of a series of speckles, rapidly varying in time and space, such that they appear smooth over plasma response times and hydrodynamic length scales. Kinetic studies, neglecting magnetic fields, have been performed examining the degree of thermal smoothing for both single mode perturbations \cite{Epperlein1988b} and optically-smoothed beam like perturbations \cite{Williams1991, Keskinen2009,Keskinen2010}. Full-physics hydrodynamic simulations of ICF targets subject to nonuniform irradiation have also been performed in two and three dimensions \cite{Demchenko2015a,Igumenshchev2016,Igumenshchev2017} and including reduced models of nonlocal effects \cite{Hu2016,Marocchino2014}. The solid density target response to pressure perturbations meanwhile is studied in \cite{Ishizaki1997,Goncharov2000,Gotchev2006}. The effects of magnetic fields on smoothing of single mode perturbations has also been studied by Bell et al.\cite{Bell1986} and Sanz et al. \cite{Sanz1988}  within a linearised hydrodynamic framework. Self-magnetisation of individual speckles has been predicted \cite{Dubroca2004,Thomas2009} and the collective magnetic field effects of a time evolving pattern of many speckles has been investigated with a reduced Braginskii transport model \cite{Rahman1997} but has not been studied  kinetically until now.
	
\begin{figure}
		\includegraphics[width=\columnwidth]{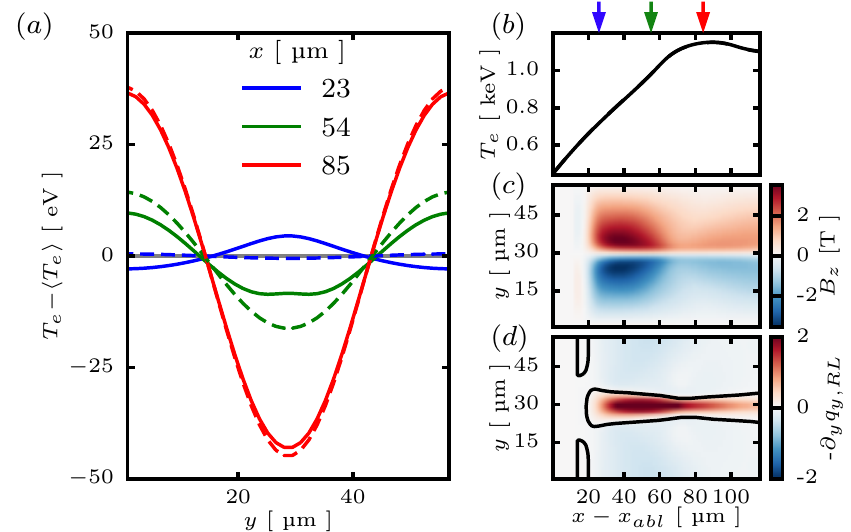}
		\caption{\label{fig:modeinversion}\label{fig:modeinversion-dyqy} Results for nonuniform laser illumination with a single mode, after \SI{372}{ps}. (a) Lateral temperature perturbation profiles  at different depths into the conduction zone, for simulations with (solid lines) and without (dashed lines) magnetic field. The horizontal grey line is included to indicate $T_e - \langle T_e \rangle = 0$. (b) Temperature profile across the conduction zone. The arrows indicate $x-x_{abl}$ positions from which the perturbation amplitude is sampled. (c) Map of conduction zone magnetic field. (d) Transverse gradient of the Righi-Leduc heat flow ($\SI{10}{eV/ps}$), the dark region indicates Righi-Leduc induced heating causing the mode inversion. }
	\end{figure}

	In this letter we aim to investigate the effect of magnetic fields and ablating ions on the degree of thermal smoothing. 2D kinetic simulations of a planar ablating target irradiated by a perturbed laser drive are carried out with the fully implicit Vlasov-Fokker-Planck (VFP) code, IMPACT  \cite{Ridgers2008a,Kingham2004a}. Two different types of heating perturbation are applied, a static single mode perturbation, $I(y) = I_0(1 + e^{i2\pi y/\lambda_P})$, and a dynamically evolving speckle pattern. The speckle pattern was designed to take the form of an SSD smoothed laser beam with a correlation time of $\SI{5}{ps}$ and a speckle width of $\SI{5}{\micro\meter}$. Even for the small Hall parameters observed ($\omega \tau_{ei}\lesssim 0.05$), for an ablating plasma subject to a single mode heating perturbation self-generated magnetic fields have a significant effect on both the fluid and heat flow dynamics within the conduction zone. The magnetic fields cause an inversion and enhancement of the temperature perturbation amplitude for the single mode perturbation,  displayed in Fig.  \ref{fig:modeinversion}, where solid and dashed lines distinguish between simulations including and omitting magnetic fields.  The inversion of the temperature perturbation, $\delta T_e = T_e - \langle T_e \rangle$, results in pressure perturbations, $\delta P = \langle n_e \rangle \delta T_e + \langle T_e \rangle \delta n_e \approx  \langle n_e \rangle \delta T_e$, of the same form.  This inversion occurs, regardless of the hydrodynamic response of the plasma, and is distinct from perturbation oscillations that can occur as a result of dynamic overpressure stabilisation of the Rayleigh-Taylor instability at the ablation front \cite{Nuckolls1972a}. In the speckle pattern simulation, magnetic fields also increase the degree of nonuniformity at the ablation surface, resulting in up to an order of magnitude reduction in the degree of thermal smoothing across the conduction zone.

	IMPACT solves the electron VFP equation,
	\begin{align}
	&\left[\frac{\partial}{\partial t}+ \mathbf{v}\cdot \frac{\partial}{\partial \mathbf{x}} + \frac{e}{m_e}\left( \mathbf{E} +  \mathbf{v} \times  \mathbf{B} \right ) \cdot  \frac{\partial}{\partial \mathbf{v}} \right] f( \mathbf{v},\mathbf{r},t) =
	 \nonumber\\ 
	 -  &\frac{\partial}{\partial \mathbf{v}} \cdot \left\{ f( \mathbf{v},\mathbf{r},t) \langle \Delta \mathbf{v} \rangle \right\} + \frac{1}{2} \frac{\partial}{\partial \mathbf{v}} \frac{\partial}{\partial \mathbf{v}}:  \left\{ f( \mathbf{v},\mathbf{r},t) \langle \Delta \mathbf{v} \Delta \mathbf{v}\rangle \right\}, \label{eq:VFP}
	\end{align}

	 in two Cartesian spatial dimensions and three velocity space dimensions using implicit finite-difference methods. Faraday's  and Amp\`ere's laws are solved self-consistently to obtain the electromagnetic fields, $(E_x, E_y,0)$ and $(0,0,B_z)$, while a magneto-hydrodynamic momentum equation is used to model the cold ions. The standard Cartesian tensor expansion \cite{Shkarovsky1966} is used  to expand the electron distribution function, $f(\myvec{v},\myvec{r},t) = f_0(v,\myvec{r},t) + \myvec{f}_1(v,\myvec{r},t) \cdot \hat{\mathbf{v}} + \dots$ (where $\hat{\mathbf{v}}$ is the velocity unit vector), in increasing degrees of velocity anisotropy. Collisions increasingly smooth out higher order terms in this expansion, $f_0 \gg \myvec{f}_1 \gg \dbul{\mathbf{f}}_2$ etc. The plasmas considered here are sufficiently collisional that we can truncate this expansion at $\myvec{f}_1$, with the error $\mathcal{O}(\lambda_{mfp}/L_T)\lesssim 2\%$. 
	
\begin{figure}
	\includegraphics[width=0.8\columnwidth]{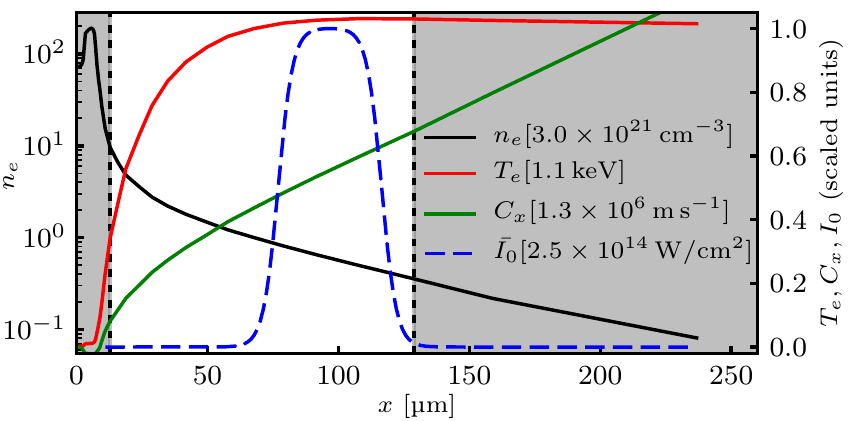}
	\caption{\label{fig:initialconditions} Initial temperature, $T_e$, number density, $n_e$ and ion velocity profiles, $C_x$, used to initiate the IMPACT simulation, between the vertical dashed black lines. $\bar{I}_0$ is the laser absorption profile. Profiles are taken from \SI{0.25}{ns} into a 1D radiation hydrodynamics simulation using the code HELIOS \cite{MacFarlane2006a}.  }
	\end{figure}

	
	A 1D radiation-hydrodynamics simulation  using the code HELIOS \cite{MacFarlane2006a} of a planar ablating foil, with mean atomic number $\langle Z\rangle = 6.51$, was performed to simulate the earliest stages of ablation, in which ionisation and radiation transport physics are important. Profiles for the ion velocity, $C_x$, electron number density, $n_e$ and electron temperature, $T_e$, taken from this HELIOS simulation, displayed in \sref{fig:initialconditions}, were used as initial conditions for the 2D IMPACT simulation. 
	
	Ion outflows, at upwards of $\approx$ \SI{100}{\kilo\meter\per\second}, are a key characteristic of ablating plasmas. These outflows are critical in correctly modelling the magnetic field dynamics within the conduction zone. B-field advection is a balance between frozen-in flow with the ions, the Nernst advection with the heat flux, and advection down resistivity gradients. The ablating plasma flows also alter the net energy flux, modifying the enthalpic heat flow, however this proves to be a less important factor. In order to include ablation, inflow and outflow boundary conditions were implemented in the code. At the inflow boundary, electrons are assumed to be in thermodynamic equilibrium; the isotropic part of the distribution function, $f_0$, is  forced to a Maxwellian with a constant number density and electron temperature. Bulk plasma flow velocity normal to this boundary is set such that $\partial_x (n_e C_x) = \mbox{constant}$. The inflow boundary is assumed unmagnetised. At the outflow, linear extrapolations into the halo cell were used for $C_x$ and $B_z$, while $f_0$ was extrapolated quadratically. An additional region of steady-state flow was added to the coronal plasma to ensure the outflow boundary did not impinge on the conduction zone physics.  An inverse bremsstrahlung heating operator \cite{Langdon1980} was used to model the perturbed laser drive, with mean intensity, $\bar{I}_0 = $ \SI{2.5e14}{\watt\per\centi\meter^2} and laser wavelength, $\lambda_L$ = \SI{351}{\nano\meter}.  Periodic boundary conditions were used in the transverse direction. The velocity grid had a cell size of $\Delta v = 0.1\,\si{v_{th}}$, and extended up to the maximum value, $\si{v_{max}}=10\,\si{v_{th}}$, where $\si{v_{th}}$ is the thermal velocity $\si{v_{th}} = \sqrt{2k_B T_{e,0}/m_e}$, evaluated at the reference temperature, $T_{e,0}= \SI{1.06}{keV}$. The longitudinal grid cell size was $\Delta x = 1.1\,\si{\micro\meter}$ within the conduction zone, while the time step was $\Delta t = 0.04\,(0.07)\, \si{ps}$, where terms in brackets indicate parameters for the speckle simulations. The transverse cell size was $\Delta y = 1.3\, \si{\micro\meter}$ for the single mode case (40 grid points per perturbation wavelength) and $\Delta y = 0.3\, \si{\micro\meter}$  for the speckle simulations.

	\begin{figure}
	\includegraphics[width=0.9\columnwidth]{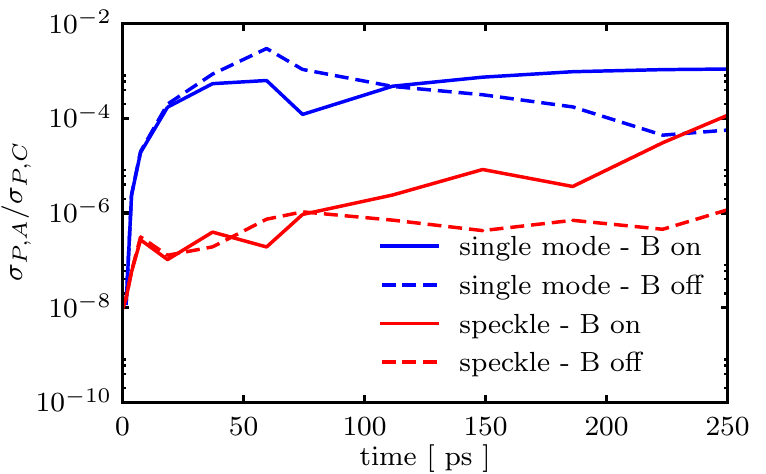}
	\caption{\label{fig:degreesmoothingasafctoftime} Evolution of the degree of smoothing across the conduction zone. The ratio of the pressure nonuniformity at the ablation and critical surfaces is shown as a function of time for a static single mode \SI{55}{\micro\meter} perturbation and a dynamically evolving speckle pattern, $\tau_c = $\SI{5}{\pico\second}, with and without magnetic fields included.}
	\end{figure}
	The integrated smoothness of the ablation pressure, $\sigma_P$, introduced by Epperlein \cite{Epperlein1988b} is defined as,
	\begin{equation}
	\sigma_{P,rms} = \left\{ \left[\int dy (P - \langle P\rangle)^2\right]/\int dy \right\}^{1/2} \langle P \rangle^{-1}.
	\end{equation}

	The ratio of this parameter's values at critical, $\sigma_{P,C}$ and ablation surfaces, $\sigma_{P,A}$, as a function of time for a $\lambda_P=$  \SI{55}{\mu\meter} wavelength perturbation are displayed in  Fig. \ref{fig:degreesmoothingasafctoftime}. Magnetic fields assist smoothing before $\approx$ \SI{100}{\pico\second} in the static single mode case but have a detrimental effect on perturbations afterwards. For the dynamically evolving speckle pattern, the detrimental effect of the magnetic field sets in at an earlier time.
	
\begin{figure}
	\includegraphics[width=\columnwidth]{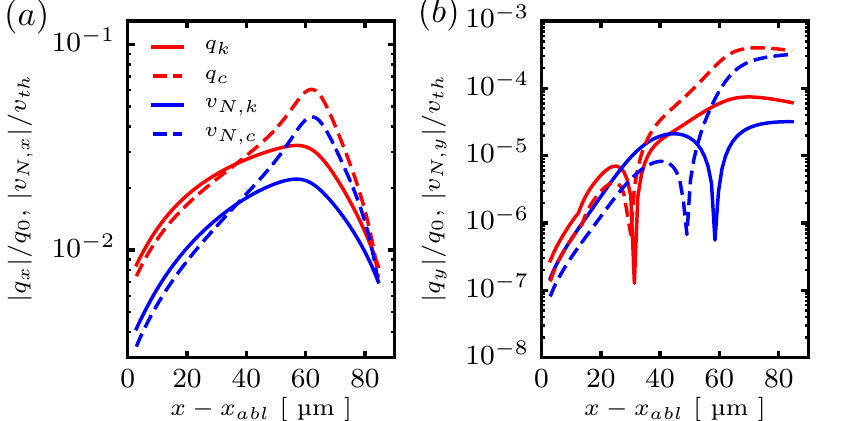}
	\caption{\label{fig:kc_comparison} Comparison of kinetic (solid lines) and classical (dashed lines) heat flow ($q$) and Nernst velocity ($v_N$) components, across the conduction zone. Results are shown for a single mode 100\% perturbed, $\lambda_P =$ \SI{55}{\micro\meter} simulation, taken from a slice normal to the target. (a) displays the $x$ component,  parallel to the bulk temperature gradient, (b) displays the transverse Nernst and heat flow components. The cross section was sampled from a transverse position \SI{12}{\micro\meter} into the simulation at a time of \SI{223}{\ps}. $q_0=\SI{1.9e15}{W/cm^2}$. }
	\end{figure}

	When B-field is included, the temperature perturbation inverts and grows through the weakly magnetised region. This can be seen in Fig. \ref{fig:modeinversion}, in which the temperature perturbation amplitude is plotted for a selection of transverse slices along the temperature gradient  for simulations with and without magnetic field after \SI{372}{\pico\second}.  $\delta T_e$ develops at the critical surface (due to modulated heating), this generates a modulated magnetic field through the Biermann battery mechanism. This magnetic field is continuously advected into the conduction zone  and amplified by the Nernst effect \cite{Nishiguchi1985}. The conduction zone  magnetic field generates an additional lateral heat flux towards the perturbation trough; the Righi-Leduc heat flow ($q_{y,RL}$). Once the B-field has developed significantly, sufficient energy is redirected by $q_{y,RL}$ that $\delta T_e$ inverts and grows.  This is clearly seen upon examination of the $q_{y,RL}$ contribution towards $\nabla\cdot \myvec{q}\propto - \partial_t T_e$.  $\partial_y q_{y,RL}$ is large and negative in the centre indicating heating at the would-be perturbation trough, \sref{fig:modeinversion-dyqy}{(d)}. Since the transport equations are not directly solved by a VFP code, the kinetic $q_{RL}$ and $v_N$ (Fig. \ref{fig:kc_comparison}) have been reconstructed a priori from the distribution function. The derivation of the kinetic analogues of the classical Ohm's law and heat flow equation \cite{Williams2013,Luciani1985}, that reproduce the correct classical expressions in the limit that $f_0$ tends to a Maxwellian, is presented in the Supplemental Material \cite{SupplementalMaterial}.
	
	
	Mode inversion occurs regardless of amplitude modulation size. Inversion is exhibited for laser profile modulations down to the 1\% level (the smallest tried). Inversion also occurs with no ion flow, when $f_0$ is forced to a Maxwellian (removing non-local effects), in a $\langle Z \rangle=3.5$ simulation, and for a broad selection of perturbation wavelengths, $\lambda_p = 10-\SI{140}{\micro\meter}$. Both B-field and temperature perturbation amplitudes are proportional to the degree of heating nonuniformity at the critical surface. Since a smaller temperature perturbation requires a proportionally smaller B-field modification of $q_y$ to achieve inversion, the mechanism is amplitude independent. For large modulations the lateral Nernst advection becomes important, compressing the B-field into the centre, increasing its peak value.
	\begin{figure}
	\includegraphics[width=\columnwidth]{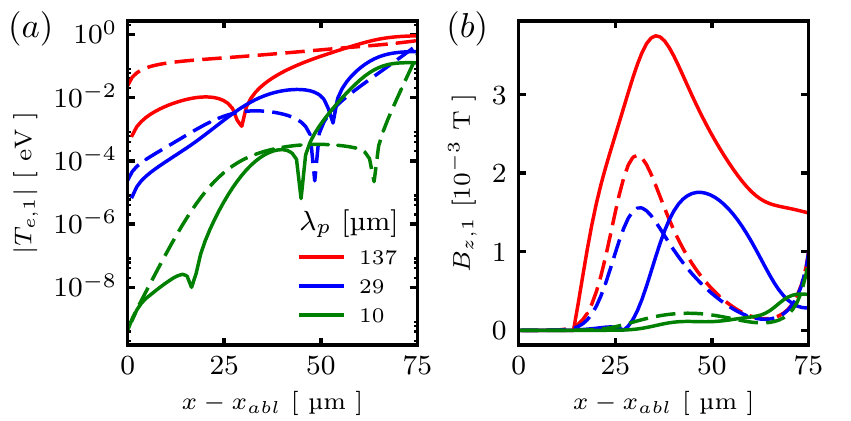}
	\caption{\label{fig:model_num+sim}\label{fig:model_num+sim-Bz} Temperature and magnetic field perturbation amplitudes across the conduction zone. Numerical solution of the reduced model (dashed lines) compared with simulation results (solid lines) for a selection of different perturbation wavelengths, $\lambda_P$, at the time of \SI{223}{\pico\second}. The cusp in $T_{e,1}$ is the position of mode inversion.}
	\end{figure}
	
	A reduced mathematical model can be used to describe the mode inversion in the linear regime. Starting with the electron temperature equation and the induction equation, we assume perturbations of the form, $T = T_0(x) + T_1(x) e^{iky}, B_z =  B_{z1}(x) e^{iky}$. Perturbations in ion velocity and number density are neglected, and the time evolution of temperature is assumed negligible compared to the B-field evolution. To first order, the linearised temperature and induction equations are, 
	
	\begin{subequations}
	\label{eq:linearised-eq}
	\begin{eqnarray}
	&(\partial_x^2& - k^2)(T_0^{5/2}T_1) = -\frac{\kappa_{\wedge,0}}{\kappa_{\perp,0}}ik \theta_0 T_0^{5/2}B_{z,1} \label{eq:linearised-eq-Te}\\
	\partial_t B_{z,1} &=& a_1\partial^2_x B_{z,1}
	+a_2 \partial_x B_{z,1} +a_3 B_{z,1} +b_1 T_{1}\label{eq:linearised-eq-Bz}\\
	\mbox{where } & a_1 =& \alpha_0, \quad
	a_2 =\left[ - C_{x,0} + \beta_{\wedge,0}  \theta_0 -\frac{3}{2}\alpha_0 \frac{\partial_x T_0}{T_0} \right ],\nonumber\\
	a_3 &=& \left[ -\partial_x C_{x,0} - \alpha_0 k^2+ \beta_{\wedge,0} \partial_x \theta_0 \right], \nonumber\\
	b_1 &=& - \frac{ik \partial_x n_0}{n_0}, \mbox{ } \alpha_0 = \frac{\alpha_{\perp,0} \delta_c^2}{T_0^{3/2}}, \mbox{and } \theta_0 = \tau_{ei,0}\partial_x T_{0}. \nonumber
	\end{eqnarray}
	\end{subequations}
	
	$\kappa_{\perp,0}$, $\kappa_{\wedge,0}$, $\alpha_{\perp,0}$, and $\beta_{\wedge,0}$ are the dimensionless diffusive thermal conduction, Righi-Leduc heat flow, resistivity and Nernst transport coefficients, respectively \cite{Epperlein1986f}. On the right hand side of Eq. \ref{eq:linearised-eq-Bz}, $a_1$ and $a_2$ represent resistive  B-field diffusion and advection respectively. $a_3$ contains the Nernst amplification, resistive and hydrodynamic damping of the B-field, while $b_1T_1$ is the Biermann battery source. $\tau_{ei,0}$ is the electron-ion collision time and is a function of $n_0(x)$ and $T_0(x)$. $\delta_c$ is the normalised collisionless skin depth, $\delta_c = c/\sqrt{n_{e,ref} e^2/\epsilon_0 m_e}$, and the asymptotic forms of the transport coefficients for small Hall parameters have been used \cite{Epperlein1984a}. Zeroth order profiles are taken from the IMPACT simulations. Fig. \ref{fig:model_num+sim}, shows numerical solutions to the above equations (dashed lines) alongside the IMPACT perturbation amplitudes (solid lines) as a function of distance from the ablation surface, for a 1\% laser intensity perturbation after \SI{223}{\pico\second}. The boundary conditions used for Eq. \ref{eq:linearised-eq} are, $B_{z,1}(x_{crit},t)=\langle B_{z,crit,VFP}\rangle$, $B_{z,1}(x,t=0)=0$ and, $T_{1,crit}=\langle T_{1,crit,VFP}\rangle$, where  $\langle B_{z,crit,VFP}\rangle$ and $\langle T_{1,crit,VFP}\rangle$ are the time averaged values of the perturbation amplitudes at the critical surface in IMPACT.
	
	The Nernst coefficient in the model has been suppressed by 35\%, inline with the average reduction in the simulation results, which are compared with Braginskii predictions in Fig. \ref{fig:kc_comparison}. The Nernst advection arises as a consequence of the magnetic field being effectively \emph{frozen} to the hot, relatively collisionless electrons, while collisions enable cold populations to diffuse more easily across field lines. In a similar manner to non-local heat flux suppression,  the depletion of the $f_0$ hot electron  tail at the top of the temperature gradient results in a reduction in the Nernst term. This results in a 4\textemdash5 fold reduction in peak B-field and brings the model into closer agreement with simulations in both amplitude and progression of B-field into the conduction zone. Kinetic modifications in the longitudinal $\kappa_{\perp}$ and lateral $\kappa_{\wedge}$ components are both approximately proportional to the total $q_x$ deviation, that can be inferred from \sref{fig:kc_comparison}{(a)}. $\kappa_{\perp}$ and lateral $\kappa_{\wedge}$ nonlocal effects, therefore, approximately cancel on the right hand side of Eq. \ref{eq:linearised-eq-Te} to not significantly change the mode inversion threshold.
	
	
	
	Qualitatively, the Nernst effect and heat flow deviate from the classical case in a similar fashion,  as shown in Fig. \ref{fig:kc_comparison}; a result consistent with the findings of Brodrick et al. \cite{Brodrick2018}.  The deviation of the Nernst velocity from its classical value tends to be more severe than for the heat flow.  The peak longitudinal suppressions are ~60\% for the Nernst and ~50\% for the heat flow. The transverse heat flux is suppressed more severely than its longitudinal component at the top of the heat front \cite{Epperlein1988b} and this is also the case for the transverse Nernst term. Hot electrons, with relatively long mean free paths, stream down the temperature gradient and preheat the cold dense plasma. This results in both heat flow and Nernst velocity values greater than classical predictions at the foot of the heat front. The ratios between kinetic and classical calculations, subscripts $k$ and  $c$, take values between $-2\leq \log|v_{N,y,k}/v_{N,y,c}|\leq 1.5$ and $-1.6\leq \log|q_{y,k}/q_{y,c}|\leq 0.8$, within the conduction zone. 
	
	
	
	\begin{figure}
	\includegraphics[width=\columnwidth]{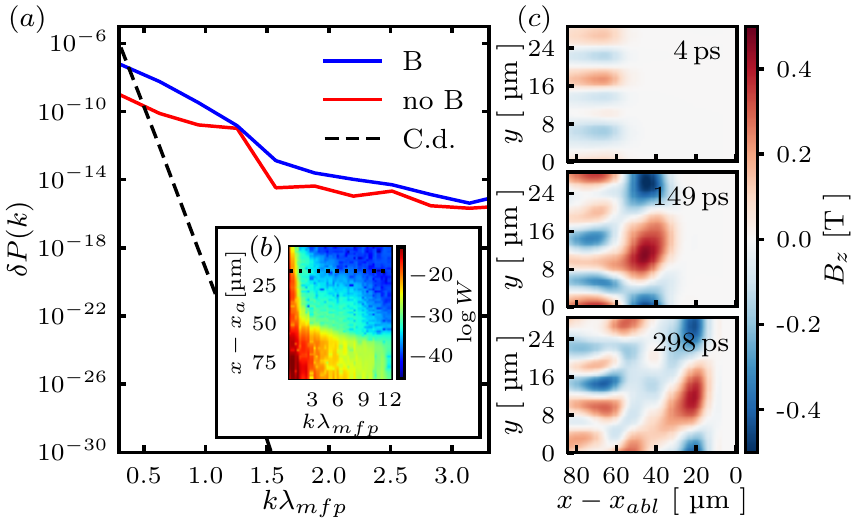}
	\caption{\label{fig:speckleB-field}\label{fig:specklesmoothing}  Results for laser illumination with dynamic speckles. (a) Pressure perturbation amplitude, $\delta P$, as a function of wave number, $k$, for a run with and without magnetic fields, taken from a cross section $\SI{20}{\micro\meter}$ from the ablation surface at \SI{298}{\pico\second}. The cloudy day model \cite{Brueckner1974} (C.d.)  is also provided for comparison. (b) The difference in perturbation amplitudes between a simulation with and without magnetic fields, $W = |\delta P_B(k)| - |\delta P_{no B}(k)|$, as a function of distance from the ablation surface and wave number, at \SI{298}{\pico\second}. (c) Time evolution of the magnetic field. The speckle coherence time and typical radius are $\tau_c$ = \SI{5}{\pico\second} and \SI{5}{\micro\meter}, respectively.}
	\end{figure}

	\sref{fig:speckleB-field}{(c)} demonstrates the time evolution of the magnetic field for the simulation in which the heating is perturbed to mimic a dynamically evolving speckle pattern. A random set of electric field amplitudes, obeying Gaussian statistics, were generated at each time step then Fourier windowed in space and time domains \cite{Feugeas2008b}. The speckle correlation time was set at \SI{5}{\ps} and the typical speckle radius is \SI{5}{\micro\meter}. 
	
	A time history of the laser nonuniformity is effectively \emph{frozen} into the conduction zone by the magnetic field, \sref{fig:speckleB-field}{(c)}. Consequently, the degree of pressure nonuniformity for the time evolving speckle perturbation continues to rise steeply at late times, \sref{fig:degreesmoothingasafctoftime}.	The Fourier spectrum of the $\delta P(y)$ near the ablation surface is compared  alongside the cloudy day model (C.d.) in \sref{fig:speckleB-field}{(a)}. The critical surface $\delta P(y)$ is time dependent, we have therefore used the time averaged critical surface pressure amplitude, $\langle \delta P(k)\rangle = \frac{1}{t_m}\int_{0}^{t_m}\delta P(k,t)dt$, in our definition of the  cloudy day model, $\delta P_{c.d.}(k) = \langle \delta P_{crit}(k) \rangle e^{-kx}$. In these VFP simulations, thermal smoothing of the high wave number components of the speckle--induced pressure perturbation is markedly less than predicted by the cloudy day model (C.d.). This is true with or without magnetic fields. However, what magnetic fields do is preferentially enhance lower wave number perturbations reaching the ablation surface. The reason for this is two fold,  high-$k$ $\delta B_{z}$ modes are preferentially damped by resistive diffusion ($\propto k^2$), and experience reduced Biermann battery growth (since the source rate $\approx \frac{T_{1,crit} \partial_x n_0}{n_0} k e^{-kx}$ extinguishes rapidly beyond the critical surface). This effect is also observed in the static single mode simulations, in which longer wavelength perturbations exhibit substantially higher B-fields, \sref{fig:model_num+sim-Bz}{(b)}. It is therefore concluded that, although the earliest times are not simulated here, the mechanism presented may lengthen the decoupling time \cite{Goncharov2000} of medium to longer wavelength modes, $k \lesssim \sqrt{(\beta_{\wedge,0} \partial_x{(\tau_{B,crit}\partial_x T_{crit})} T_{crit}^{3/2} )/(\alpha_{\perp,0}\delta_c^2)}$.
	
	In summary, 2D kinetic simulations of a planar-ablating foil have been performed, including, for the first time, both magnetic field effects and realistic ablating outflows. Once enough time has passed for the magnetic field to be advected into the conduction zone and amplified, the magnetic field enhances pressure perturbation amplitudes in both the case of a single mode perturbation and a time evolving speckle pattern. Even for the small Hall parameters seen here, the transverse Righi-Leduc heat flow is on the order of the transverse diffusive heat flow and is sufficient to cause an inversion of a static single mode perturbation and to distort the heat front. This mode inversion mechanism is robust, occuring over a wide range of laser non-uniformity amplitude. Magnetic fields are more detrimental to lower wave number perturbations as they are less susceptible to resistive damping. The effects of changing speckle pattern coherence times, different plasma regimes and how magnetic fields will interface with hydrodynamic instabilities at the ablation surface, will be the subject of further work.  This work highlights that both self-generated magnetic fields and kinetic effects are required in order to correctly model the conduction zone in direct-drive ICF, and to precisely calculate thermal smoothing in particular. The mechanism presented may alter the required tolerances for beam nonuniformity in ICF implosions and could be measured by experiment.

	\begin{acknowledgments}
	We thank K. McGlinchey, H. Watkins, S. Mijin and S. Hooper-Kay for useful discussions. Simulations results were obtained with the use of the Imperial College Research Computing Service  \cite{HPC}.  This work was supported by the Engineering and Physical Sciences Research Council through Grant No. EP/J500239/1.
	\end{acknowledgments}
	
	\bibliography{library}
	
	\end{document}